\documentclass[12pt,preprint]{aastex}

\shorttitle{Formation and evolution of G1}
\shortauthors{Baumgardt et al.}

\begin{document}


\title{A Dynamical Model for the Globular Cluster G1}


\author{Holger Baumgardt\altaffilmark{1},        
        Junichiro Makino\altaffilmark{1},
        Piet Hut\altaffilmark{2},
        Steve McMillan\altaffilmark{3},
        Simon Portegies Zwart\altaffilmark{4}}

\altaffiltext{1}{
        Department of Astronomy, University of Tokyo, 7-3-1 Hongo,
        Bunkyo-ku,Tokyo 113-0033, Japan}

\altaffiltext{2}{
        Institute for Advanced Study, Princeton, NJ 08540, USA}

\altaffiltext{3}{
        Department of Physics, Drexel University, Philadelphia, PA
        19104, USA}

\altaffiltext{4}{
        Astronomical Institute ``Anton Pannekoek'' and Section Computational Science, University of
        Amsterdam, Kruislaan 403, 1098 SH Amsterdam, The Netherlands}


\begin{abstract}

We present a comparison between the observational data on the
kinematical structure of G1 in M31, obtained with the Hubble WFPC2 and
STIS instruments, and the results of dynamical simulations carried out
using the special-purpose computer GRAPE-6.  We have obtained good
fits for models starting from single cluster King-model initial conditions
and even better fits when starting our simulations with a dynamically
constructed merger product of two star clusters.
In the latter case, the results from our simulations are in excellent
agreement with the observed profiles of luminosity, velocity dispersion,
rotation, and ellipticity.  We obtain a
mass-to-light ratio of $M/L = 4.0 \pm 0.4$ and a total cluster mass of
$M=(8\pm 1)\times 10^6 \; M_\odot$.  
Given that our dynamical model can fit all available observational
data very well, there seems to be no need to invoke the presence of an
intermediate-mass black hole in the center of G1.

\end{abstract}




\newcommand{\msun}{M_{\odot}}
\def\apgt{\ {\raise-.5ex\hbox{$\buildrel>\over\sim$}}\ }
\def\aplt{\ {\raise-.5ex\hbox{$\buildrel<\over\sim$}}\ }

\section{Introduction}

We report results from a series of $N$-body simulations for the
globular cluster G1 in M31.  G1 is one of the brightest and most
massive globular clusters in the local group. Its total luminosity
($M_V=-10.94$ mag) and  central velocity dispersion
($\sigma_0 = 25.1 \pm 1.7$ km/sec) are larger than those of
any galactic globular cluster \citep{Meylanetal2001, Djorgovskietal1997}.

\citet{Meylanetal2001} used virial mass estimates and mass estimates from 
King-Michie models for G1.  They obtained total masses in the range 
$7.3 \times 10^6 M_{\odot}$ to $15\times 10^6 M_{\odot}$ and (for their model 4) 
a core radius, 
half-mass radius, and tidal radius of 0.53, 13.2, and 187 pc, respectively. 
The estimated half-mass relaxation time was 50 Gyr, much longer than the Hubble
time.

\citet{Gebhardtetal2002} have reported evidence for an intermediate-mass
black hole of $2.0^{+1.4}_{-0.8}\times 10^4 M_{\odot}$ in the center
of G1.  Based on velocity profiles obtained with the Hubble space
telescope's STIS instrument, they constructed orbit-based axisymmetric models.
Varying $M/L$ and the mass of the central black hole, they found
a best fit for $M/L=2.5$ and $M_{BH}=2\times 10^4 M_{\odot}$.
A model without a central black hole was rejected at a $2\sigma$ level.

The presence of such a black hole would be very interesting, for at
least two reasons.  First, it would lie neatly on the  $M_{BH} - \sigma$
relation for galaxies \citep{Gebhardtetal2000,FerrareseMerritt2000}. 
Second, G1 would then be a good example of the type of cluster postulated by 
\citet{Ebisuzakietal2001}, some of which may find their way into the
center of a galaxy by dynamical friction, where their intermediate-mass
black holes may then merge to provide the seeds for supermassive
black holes.

However, before embracing such an exciting conclusion it is all the more
important to ensure that more conventional explanations of the
observational data are ruled out.  To this end, we have tried to
construct the best possible evolutionary model for G1 as a large
globular cluster that is still in the early stages of core collapse,
without harboring an intermediate-mass black hole.  We have run a set of 
models with varying
initial density profiles, half mass radii, total masses, and global $M/L$
until we found a model that gave the best fit to the light
and velocity profiles of G1.

In \S2 we describe our numerical method.  In \S3 we present the results
of simulations starting with a single non-rotating cluster, and in
\S4 we show what happens when we consider G1 to be the rotating
product of a merger of two smaller globular clusters.  We briefly
summarize in \S5.

\section{Modeling Method}

In order to model the evolution of G1 using $N$-body simulations, we
face a scaling and a fitting problem: we can only handle $\sim 10^5$
particles while G1 contains $\sim 10^7$ stars; and we do not know
which values to assign to the initial cluster model parameters such as
the total mass and the half-mass radius. We solve
the scaling problem by scaling the dynamical parameters in such as way
as to reproduce in our model simulations the correct two-body
relaxation time scales inferred for G1 from observations.  We solve the
fitting problem by carrying out a large enough number of runs to allow
us to isolate simulations that closely reproduce the observational data.
Without the use of the GRAPE-6 computers \citep{Makinoetal2003}, it
would have been unpractical to run the several dozen runs needed to
determine our best fits.

We used Aarseth's $N$-body code NBODY4 \citep{Aarseth1999}.  All
simulated clusters contained $N=65,536$ stars initially, with a range
of masses following \citet{Kroupa2001}'s mass function with lower and
upper mass limits of 0.1 and 30 $\mbox{M}_\odot$, respectively.  Our
simulations did not contain primordial binaries, which is a reasonable
simplification for a cluster that is still quite far from core
collapse.  We did not include M31's galactic tidal field, which would
have a negligible influence at the position of G1, at least 40 kpc
from the center of M31. Since tidal effects are unimportant,
we are left with two evolution mechanisms: stellar evolution and 
two-body relaxation.

Stellar evolution was modeled according to the fitting formulae of
\citet{Hurleyetal2000}, using a metallicity
of $\mbox{[Fe/H]} = -0.95$, similar to the mean metallicity of G1 as
determined by \citet{Meylanetal2001}.  We assumed a retention fraction
of neutron stars of 15\%.

All simulations were carried out
for 13 Gyr and the final density and velocity profiles were obtained
from 10 snapshots spanning a 500 Myr period centered at $T=12$ Gyr.  For
comparison of our models with the observations of G1, we assume a
distance of 770 kpc to M31, so one arcsecond corresponds to 3.7
pc.  Typically, about 1\% of the stars escaped from the cluster during
a simulation.  Only bound stars were used for the comparison with observations.

We have to scale the parameters of our simulations in order to match
the most important stellar evolution and stellar dynamical parameters
of the actual G1 cluster. In order to match 
the relaxation time of G1, we have to increase the radius of our cluster 
by 
\begin{equation}
     r_{h S} = r_{h G1} \; \cdot \; \left( \frac{N_{G1}}{N_S} \right)^{1/3} \left( \frac{\ln (\gamma N_S)}
{\ln (\gamma N_{G1})} \right)^{2/3} \;\; , 
\label{rscale}
\end{equation}
where subscripts $G1$ and $S$ denote, respectively, the actual values for 
G1 and those used in our simulations. Effects which 
strongly depend on the number of particles in the cluster are 
unimportant before the cluster goes into core collapse, so our models should
give a valid description of the dynamical evolution of G1 up to 
the present time. 

In the first set of simulations, we started from isotropic King model 
conditions with no initial mass-segregation,
and dimensionless central concentrations in the range $4.0 \le W_0 \le 11.0$.
For each choice of initial density profile,
we ran full simulations for a number of choices for the
initial physical half-mass radius $r_h(t=0)$ and mass $M(t=0)$ of G1
until we could fit the surface density profile of \citet{Meylanetal2001} 
over a maximum range in
radius while simultaneously obtaining an optimal fit to the observed
velocity profile.  
For the surface density, we scaled our predicted
profile by a multiplicative factor to obtain the best fit (in practice
changing the M/L predicted by our assumed IMF by a factor of $1.5 \sim 2$).
For the velocity profile, we used the symmetrized
profile shown by \citet{Gebhardtetal2002} in their Fig.~1 and the
ground-based value of \citet{Djorgovskietal1997}, who measured a
velocity of $25.1 \pm 1.7$ km/sec inside an aperture of $1\farcs15 \times 
7\farcs0$.  For each run, a best fit was determined by a $\chi^2$
test against the combined data.  With improved estimates for
$r_h(t=0)$ and $M(t=0)$, a new initial half-mass radius 
could be calculated and a new simulation was performed. Simulations 
were performed until the half-mass radius
changed by less than 5\% between successive iterations. 
A more detailed
description of our simulations and their results will be presented in
a forthcoming paper \citep{Baumgardtetal2003b}.

\section{Single Cluster Simulations}

\begin{figure}[htbp!] 
\plotone{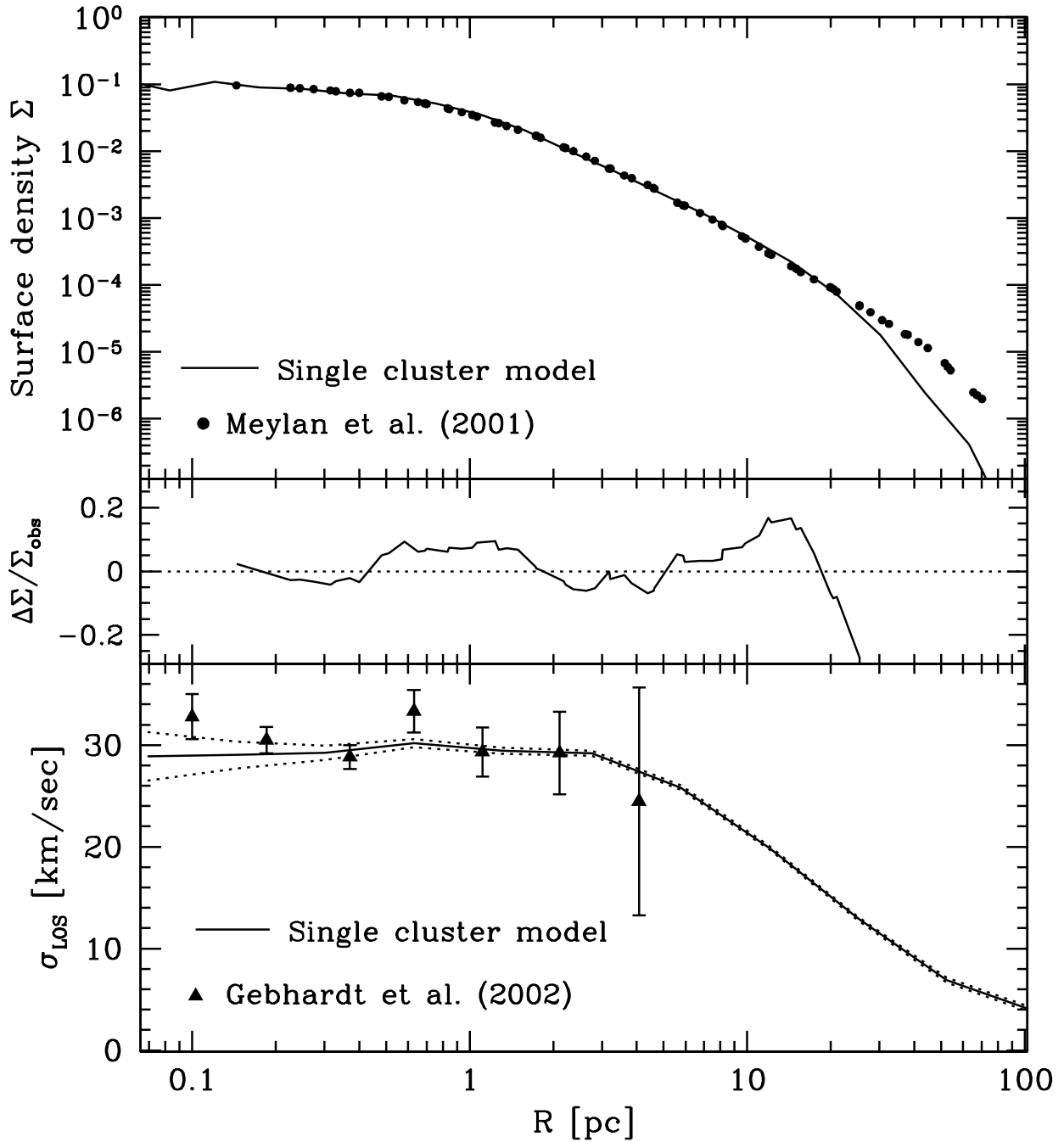}
\caption{The best-fit result, starting from a single $W_0=7.5$  King model.
The top panel shows the surface luminosity profile $\Sigma$ of 
the simulations (full line) and the observations (filled circles).
The middle panel shows $\Delta \Sigma /\Sigma_{\rm obs}$, where
$\Delta \Sigma = \Sigma_{\rm model}-\Sigma_{\rm obs}$.
The
bottom panel shows the velocity dispersion in the simulations (solid
curve, with dashed curves indicating the $1\sigma$ error) and
observations (filled triangles with error bars). 
\label{fig1}}
\end{figure}

Figure 1 shows the data for the best fit from among the runs where we
started with a single cluster in the form of a King model;
here the initial central potential depth was $W_0 = 7.5$.  The top
panel shows the inferred projected luminosity density.
We can see that the fit is very good for $r<15$ pc.  The reason
why the model density drops off sharply at large radius is because the
initial King model had a tidal radius of 32 pc if we scale it to G1.  
Two-body relaxation begins to produce an
extended halo with a surface density slope $\sim -4$, but a Hubble
time is too short to let this effect propagate very far into the observed halo.
Starting from deeper King models ($W_0=8$ or higher) does not solve
this problem: such models predict too high a surface density around
$r=10$ pc, while still falling short at larger radii.  The
implication is that G1 must have started with a density distribution
more extended than any King model that can be fit to the bulk of the stars. 

The bottom panel shows the velocity dispersion inferred from our
$W_0=7.5$ model, as compared to the dispersion observed by
\citet{Gebhardtetal2002}.  For larger $W_0$ values, our models produce
velocities that are too high at the largest observed radii.  Models
with slightly lower concentration give a somewhat better fit, but when
we require the model to reproduce the density as well, the combined
requirements clearly point to $W_0=7.5$ as producing the best agreement,
and one which falls within the observational errors everywhere except
near the tidal radius artificially imposed by the initial conditions;
we address this limitation in the next section.

Note that our model cluster has a mass smaller than those
found from multi-mass King model fits by \citet{Meylanetal2001}. Their
extreme values stem from the implicit King-model requirement that a
cluster has complete mass-segregation, which is unphysical in a massive
cluster like G1 where the relaxation time is much longer than a
Hubble time.

To sum up, an evolutionary model starting from a King model without
initial mass segregation reproduces both the luminosity profile and
the velocity dispersion profile of G1 rather well.  The fits are not
perfect, though, on two counts.  First, the best-fit model still
produces too steep a luminosity profile at larger radii. Second, since  
we start from a spherically symmetric non-rotating model, in principle
we cannot fit the observed rotation profile or ellipticity. 
The question is whether we can introduce rotation while
simultaneously at least preserving, and hopefully improving, the
reasonable fits obtained so far.  In the next section we answer this
question affirmatively.

\section{Merger Simulations}

\begin{figure}[htbp!]
\plotone{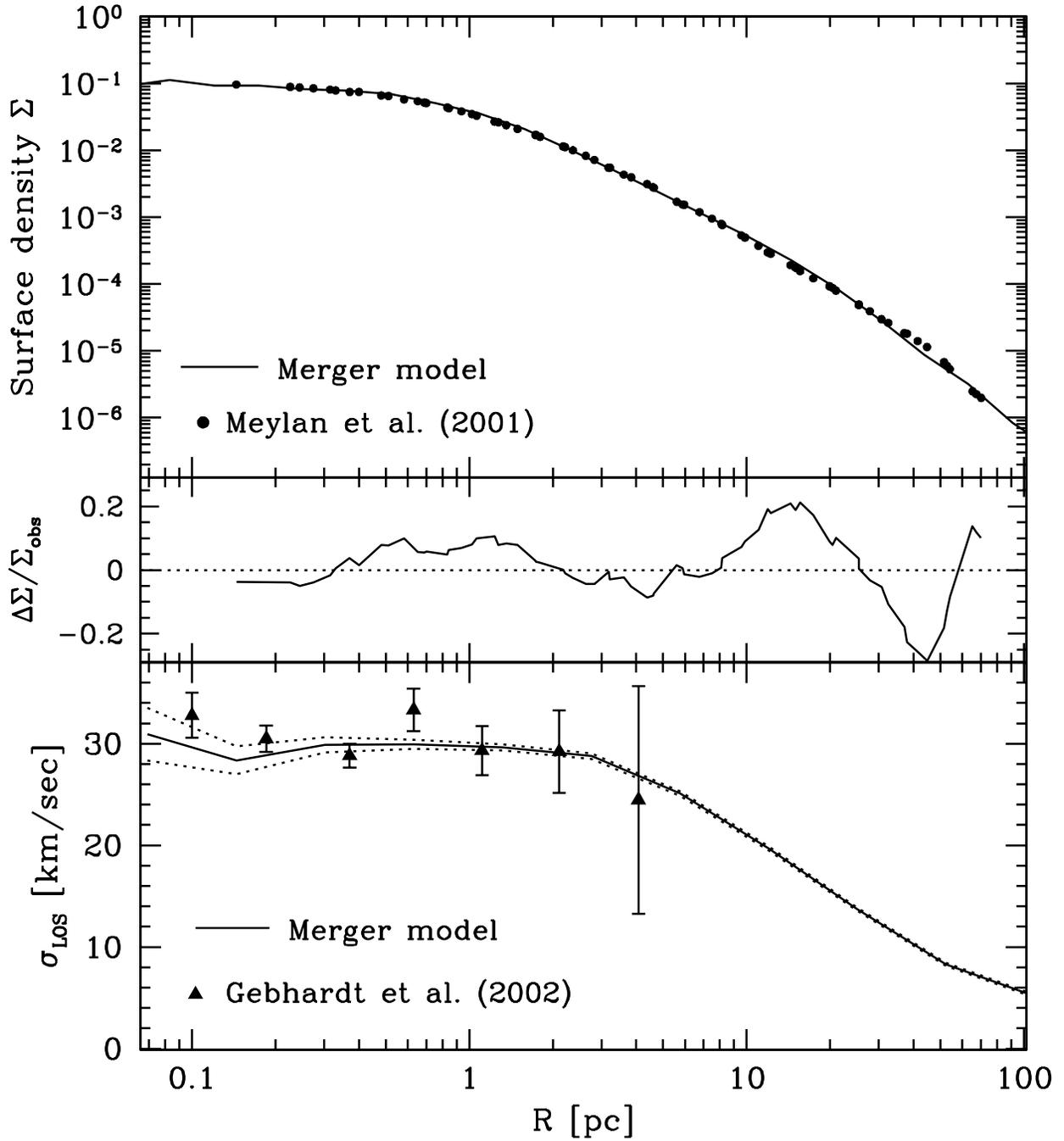}
\caption{Same as figure 1, but for the merger model that started from
two  $W_0=6.5$ King models.
\label{fig2}}
\epsscale{1.0}
\end{figure}

Currently favored scenarios for the formation of star clusters are the
collapse of giant molecular clouds or the collision of smaller clouds
\citep{FallRees1985, FujimotoKumai1997}.  A collision scenario could
easily explain the apparent rotation of G1.  It might also account for
the run of surface density in the halo, since simulations of the
merging of two stellar systems usually give surface density profiles
$\Sigma(R) \sim R^{-3.0}$ \citep{SugimotoMakino1989,Makinoetal1990,
Okumuraetal1991}.  

Based on these theoretical hints, we have carried out a series of
simulations starting with an early merger of two star clusters.  For
the sake of simplicity, we have restricted ourselves to the merger of
two identical King model clusters on parabolic orbits with initial
separation $r_i = 20$ and pericenter separation in $N$-body
units of $p = 1$ (another simulation with $p
= 2$ gave similar results). The chosen pericenter distance corresponds to
approximately 1.3 half-mass radii for the initial clusters.
We used $N=80,000$ equal-mass stars in our merger simulation without
including stellar evolution, a reasonable approximation given that our
merger was postulated to occur during formation of the clusters.
After the merger
product had undergone its violent relaxation, we randomly selected
65,536 stars from among all the stars still bound to the final merger product,
assigned masses drawn from a \citet{Kroupa2001} IMF to them,
and then started our
dynamical evolution simulations for a duration of $T=13$ Gyr.

Fig.~\ref{fig2} shows the final density and velocity dispersion
profiles for our best-fit simulations which started with a
collision between two $W_0=6.5$ initial King models.  Note that our
simulations now reproduce the observed extended halo very well.
The agreement between the observed and model
velocity dispersions is also very good.

\begin{figure}[tbp!]
\plotone{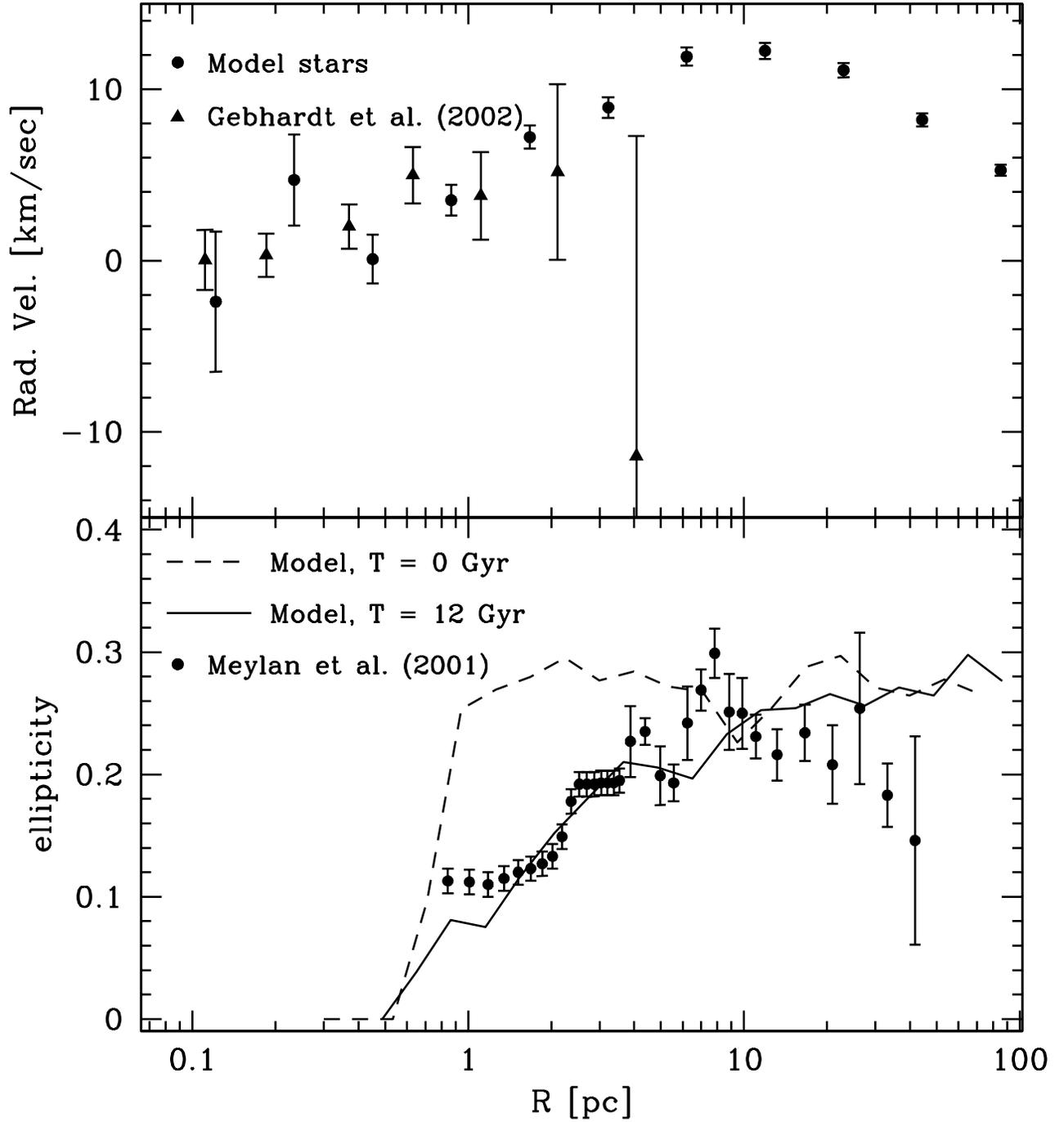}
\caption{Symmetrized radial velocities (top panel) and ellipticity
profiles (bottom panel) for the observations and our best-fit merger
model.  The ellipticity is defined as $\epsilon = 1-b/a$, where $a$
and $b$ are the major and minor axis of the best-fit ellipse for the
observations and the projected model data. 
Note that we have omitted the innermost 3 datapoints from 
\citet{Meylanetal2001}, since the ellipticity in the center 
is not well defined (Meylan 2003, private communication). 
\label{fig3}}
\end{figure}

Figure \ref{fig3} compares the rotation and ellipticity
profiles of our merger model and the observations.  We measured the
rotation profile from two directions perpendicular to each other and
the minor axis and took the mean of the two directions.  The profiles
were determined from the radial velocities of all bright stars located
in an area between angles of $10\degr$ and $40\degr$ with respect to
the major axis, in order to make an optimal comparison with
\citet{Gebhardtetal2002}, who performed their HST/STIS spectroscopy at
an angle of $25\degr$ against the major axis. The agreement 
between simulated and observed rotation profiles is very good.

\def\ltsim{\mathrel{\raise.3ex\hbox{$<$}\mkern-14mu
             \lower0.6ex\hbox{$\sim$}}}

Similarly, the ellipticity profile of our merger run is in good agreement
with the ellipticity profile of G1 as determined by \citet{Meylanetal2001},
(as can be seen in the lower panel of Fig. \ref{fig3}).
The $N$-body run starts with a near constant ellipticity of about $\epsilon=0.25$.
After $T=12$ Gyr, the cluster core has become almost 
spherical due to relaxation effects, while the halo ellipticity has remained
unchanged. The observations show a similar drop of $\epsilon$ toward the core.

Our success in modeling G1 as a merger product does not necessarily
imply that a merger history is the only way to explain its current
state.  For example, it is also possible that G1 is a heavily
stripped remnant of a dwarf spheroidal.  What is important is that
the observed rotation can be well modeled under at least one set of
reasonable assumptions, as we have shown here.  The presence of
rotation does not invalidate attempts at modeling under the simpler
assumption of spherical symmetry; rather it invites a further
fine-tuning of the already good agreement of spherical models.

Table~1 summarizes our results for the best-fitting models. It shows
$W_0$ for each initial model, as well as the half-mass radius
$r_h$ at $T=12$ Gyr, and the
inferred total mass $M$ and $M/L$ required to give the best-fit
velocity dispersions. The half-mass radius shown is the half-mass
radius of our models after they were scaled back to G1 by eq.\ 1. 
The errors given for the $M$ and $M/L$ values are the statistical errors 
from the $\chi^2$-fit. 
The global mass-to-light ratios we obtain in our best fits lie around
$M/L \sim 4$, relatively large but
still within the range of mass-to-light ratios observed for galactic
globular clusters \citep{PryorMeylan1993}.
Column 6 gives the probability $P_V$
that our velocity distribution agrees with the observations 
of \citet{Gebhardtetal2002} and \citet{Djorgovskietal1997}, 
determined from a $\chi^2$-test against the combined data.

\begin{table*}[t]
\caption[]{Results of the best-fitting $N$-body runs.}
\begin{tabular}{rccccccccc}
\noalign{\smallskip}
\multicolumn{1}{c}{$W_0$} & Type & $r_h$ & $M$ & M/L & $P_V$ & $(M/L)_C$ & $T_{RH}$ & $T_R(0)$\\
 & & [pc] & [$M_\odot$]&  & & & [Gyrs] & [Gyrs]\\
\noalign{\smallskip}
 7.5 & Single & 6.76 & $(7.60 \pm 0.76)\times 10^6$ & $3.80 \pm 0.38$ &  0.20 & 2.41 & 30.3 & 0.292\\
 6.5 & Merger & 8.21 & $(8.20 \pm 0.85)\times 10^6$ & $4.10 \pm 0.42$ &  0.28 & 2.44 & 29.5 & 0.181\\
\end{tabular}
\end{table*}

The last three columns of Table 1 give the $M/L$ values inside the core
(defined as the region containing the innermost 1\% of 
bright stars) and the half-mass and core relaxation times calculated from 
the cluster parameters at $T=12$ Gyrs and eq.\ 2-62 of \citet{Spitzer1987}. 
Since the half-mass relaxation 
time is much longer than a Hubble time, G1 has not yet reached core collapse
and the core of G1 is still dominated by low-mass main-sequence stars.
Nevertheless, core $M/L$ values are
smaller than global ones since mass-segregation has caused bright
stars, which are more massive than average, to sink into the center. 
Shortly before core collapse, bright stars will be depleted 
from the center by the heavier neutron stars and massive white dwarfs,
as in the simulations of M15 of \citet{Baumgardtetal2003a}.

\section{Conclusions}

We have constructed evolutionary models for the massive globular
cluster G1.  Starting from a $W_0=7.5$ King model we can reproduce
both the observed luminosity profile for $r<15$ pc and the
observed velocity dispersion profile.  A model starting from the
merger of two $W_0=6.5$ King models fares even better: it can
reproduce the luminosity, velocity dispersion, rotation 
profiles and ellipticity for the entire range of observations.

Our simulations were motivated by the recent claim of evidence for a
massive central black hole.  Given that our dynamical model without
central black hole can fit all available observational data very well,
there seems to be no need to invoke the presence of an intermediate-mass
black hole.  Note that we obtained an excellent fit by varying only the
following basic parameters: the central potential $W_0$, 
the initial total mass $M(0)$ and total mass-to-light ratio $M/L$, and the
initial half-mass radius $r_h(0)$.  Our conclusions are therefore
robust, and independent of any fine tuning in initial conditions.

The 2-sigma evidence presented by Gebhardt et al. (2002) for a
massive black hole is not supported by direct observation of
luminosity profile, velocity dispersion and rotation. It must have
come from the data not presented in their paper ({\it e.g.} the higher
order moments of the velocity profiles, together with multiparameter
fits to orbit families).
Without independent checks or further observational support, we
consider the evidence for such a black hole to be inconclusive.

This work is the first example of the successful detailed dynamical
modeling of the evolution of a globular cluster with rotation.  We
have shown how $N$-body simulations have matured as the
most powerful tool to interpret detailed observational data, obviating
the need for simplifying assumptions such as spherical symmetry or the
use of static ({\it e.g.} multi-mass King) models.

\section*{Acknowledgments}

The authors thank Toshi Fukushige and Yoko Funato for stimulating
discussions and Karl Gebhardt for sharing his velocity data on G1 with us. 
We are especially grateful to Sverre Aarseth for making the NBODY4 code 
available to us and his constant, untiring help with the code. This work is 
supported in part by Grant-in-Aid for
Scientific Research B (13440058) of the Ministry of Education,
Culture, Science and Technology, Japan, by NASA ATP grant
NAG5-10775 and by the Royal Dutch Academy of Science (KNAW) and Dutch
organization for Scientific Research (NWO).


\begin{thebibliography}{}
\bibitem[Aarseth(1999)]{Aarseth1999}
Aarseth, S. J. 1999, \pasp, 111, 1333
\bibitem[Baumgardt et al.(2003a)]{Baumgardtetal2003a}
Baumgardt, H., Hut, P., Makino, J., McMillan, S.L.W., and Portegies
Zwart, S.F. 2003a, \apj, 582, L21 
\bibitem[Baumgardt et al.(2003b)]{Baumgardtetal2003b}
Baumgardt, H., Makino, J., Hut, P., McMillan, S.L.W., and Portegies
Zwart, S.F. 2003b, in preparation
\bibitem[Djorgovski et al.(1997)]{Djorgovskietal1997}
Djorgovski, S. G., et al.\ 1997, \apj, 474, L19 
\bibitem[Ebisuzaki et al.(2001)]{Ebisuzakietal2001}
Ebisuzaki, T., et al.\ 2001, \apj, 562, L19 
\bibitem[Fall and Rees(1985)]{FallRees1985}
Fall, S. M., and Rees, M. J. 1985, \apj, 298, 18 
\bibitem[Ferrarese and Merritt(2000)]{FerrareseMerritt2000}
Ferrarese, L., \& Merritt, D., 2000, \apj, 539, L9
\bibitem[Fujimoto and Kumai(1997)]{FujimotoKumai1997}
Fujimoto, M., and Kumai, Y. 1997, \aj, 113, 249 
\bibitem[Gebhardt et al.(2000)]{Gebhardtetal2000}
Gebhardt, K., et al., 2000, \apj, 539, L13
\bibitem[Gebhardt et al.(2002)]{Gebhardtetal2002}
Gebhardt, K., Rich, R. M., and Ho, L. C. 2002, \apj, 578, L41
\bibitem[Hurley et al.(2000)]{Hurleyetal2000}
Hurley, J. R., Pols, O. R., and Tout, C. A. 2000, \mnras, 315, 543
\bibitem[Kroupa(2001)]{Kroupa2001}
Kroupa, P. 2001, \mnras, 322, 231
\bibitem[Makino, Akiyama, \& Sugimoto(1990)]{Makinoetal1990} Makino, 
J., Akiyama, K., \& Sugimoto, D.\ 1990, \pasj, 42, 205 
\bibitem[Makino et al.(2003)]{Makinoetal2003}
Makino, J., Fukushige, T., and Namura, K. 2003, in preparation
\bibitem[Meylan et al.(2001)]{Meylanetal2001}
Meylan, G., Sarajedini, A., Jablonka, P., Djorgovski, S. G., Bridges, T., and 
Rich, R. M. 2001, \aj, 123, 830
\bibitem[Okumura et al.(1991)]{Okumuraetal1991}
Okumura, S. K., Ebisuzaki, T., and Makino, J. 1991, \pasj, 43, 781  
\bibitem[Pryor and Meylan(1993)]{PryorMeylan1993}
Pryor C., and Meylan, G. 1993, in {\it Structure and Dynamics of Globular
Clusters}, eds.\ S.\ Djorgovski, G.\ Meylan, ASP Conf.\ Series 50,
p.\ 357
\bibitem[Spitzer(1987)]{Spitzer1987}
Spitzer, L. J. 1987, {\em Dynamical Evolution of Globular Clusters}.
 Princeton University Press, Princeton, New Jersey.
\bibitem[Sugimoto and Makino(1989)]{SugimotoMakino1989}
Sugimoto, D., and Makino, J. 1989, \pasj, 41, 1117
\end{thebibliography}
\end{document}